\documentclass{svjour3}                     % onecolumn (standard format)
\smartqed  % flush right qed marks, e.g. at end of proof
\usepackage{graphicx}
\usepackage{mathptmx}      % use Times fonts if available on your TeX system
\usepackage{natbib}
\newcommand {\gtsim} {\ {\raise-.5ex\hbox{$\buildrel>\over\sim$}}\ }
\newcommand {\ltsim} {\ {\raise-.5ex\hbox{$\buildrel<\over\sim$}}\ }

\newcommand{\cc}{cm$^{-3}$}

\newcommand{\ce}{charge exchange}
\newcommand{\kms}{km s$^{-1}$}
\newcommand{\HI}{H$^{\rm o}$}
\newcommand{\HII}{H$^{\rm +}$}

\newcommand{\nH}{$n$(\HI)}
\newcommand{\np}{$n$(\HII)}

\newcommand{\jgr}{J.~Geophys.~Res. }
\newcommand{\apj}{Astrophys.~J. }
\newcommand{\apjl}{Astrophys.~J.~Lett. }
\newcommand{\apjs}{Astrophys.~J.~Supp. }
\newcommand{\aap}{Astr.~Astrophys. }
\newcommand{\ssr}{Space~Sci.~Rev. }
\newcommand{\mnras}{MNRAS }
\newcommand{\nat}{Nature }
\journalname{SSRv}
\begin{document}

\title{The Heliosphere in Time\thanks{To appear in Space Science Reviews,
"From the Outer Heliosphere to the Local Bubble", J. Linsky et al.
(eds); ISSI workshop held Oct. 2007}}


\author{H.-R. M\"uller \and P. C. Frisch \and B. D. Fields \and G. P. Zank}
\authorrunning{M\"uller \and Frisch \and Fields \and Zank} % if too long for running head

\institute{Hans-R. M\"uller \at
              Department of Physics and Astronomy,
              Dartmouth College,
              Hanover, NH 03755, USA \\
              Tel.: +1-603-6460428\\
              \email{hans.mueller@dartmouth.edu}           %  \\
           \and
           Priscilla C. Frisch \at
              \email{frisch@oddjob.uchicago.edu}           %  \\
           \and
           Brian D. Fields \at
              \email{bdfields@uiuc.edu}           %  \\
           \and
           Gary P. Zank \at
              \email{zank@cspar.uah.edu}           %  \\
}

\date{Received: 15 April 2008 / Accepted: 30 September 2008}

\maketitle

\begin{abstract}
Because of the dynamic nature of the interstellar medium, the Sun
should have encountered a variety of different interstellar
environments in its lifetime.  As the solar wind interacts with
the surrounding interstellar medium to form a heliosphere,
different heliosphere shapes, sizes, and particle contents result
from the different environments.  Some of the large possible
interstellar parameter space (density, velocity, temperature) is
explored here with the help of global heliosphere models, and the
features in the resulting heliospheres are compared and discussed.
The heliospheric size, expressed as distance of the nose of the
heliopause to the Sun, is set by the solar wind - interstellar
pressure balance, even for extreme cases. Other heliospheric
boundary locations and neutral particle results correlate with the
interstellar parameters as well. If the H$^\circ$ clouds
identified in the Millennium Arecibo survey are typical of clouds
encountered by the Sun, then the Sun spends $\sim 99.4$\% of the
time in warm low density ISM, where the typical upwind heliosphere
radii are up to two orders of magnitude larger than at present.

\keywords{Global heliosphere modeling \and time-dependent
interstellar conditions}
\end{abstract}

\section{Introduction}\label{sec:intro}

The expansion of the coronal solar wind is terminated by
encountering the interstellar material surrounding the solar
system, the circum-heliospheric interstellar medium (CHISM). The
Sun is currently embedded in partially ionized material, with
about three fourths of the material in neutral form. Absorption
studies toward nearby stars reveal a very inhomogeneous local
interstellar medium, with many distinct velocity components in
most lines of sight. Also on bigger scales, the ISM is a dynamic
medium, constantly replenished and mixed anew by the outflows of
star forming regions and supernova explosions, for example. Shock
passages create large regions of almost empty, high-temperature
regions in their wake which may be unstable and hence short-lived.

On its journey around the Galaxy, the Sun is likely to encounter a
variety of such interstellar environments, each of which is
characterized by a set of physical quantities, including density,
temperature, magnetic field, and velocity state. As the solar wind
is often assumed to have been relatively stable over long periods
of time, the question arises how the heliosphere has reacted to
the differing interstellar environments encountered through time.
We characterize in this paper the heliospheric response to some
non-catastrophic changes in the interstellar environment
(``galactic weather''). Studies like this, paired with cosmic ray
calculations and investigations of the terrestrial consequences of
changes in the interplanetary medium due to galactic weather
events, will contribute to the understanding of long-ago events
that left their mark in terrestrial records.

A concrete example is the evidence that the relatively warm and
dense material of the contemporary local interstellar cloud (LIC)
is thought to be embedded in a larger structure, the Local Bubble
(LB). With some reasonable assumptions, it can be estimated that
the Sun was exposed to the tenuous, very hot LB environment for a
long time, has only entered the LIC some 40,000 years ago, and
will exit it in 0--4000 years from now. In this sense, roughly
steady interstellar conditions are short-lived, and changes will
occur on different time scales, with cloud passages as short as
$~10^3$ years.

The variety of possible heliosphere boundary conditions that
follow from our understanding of interstellar clouds dictate that
diverse models of the heliosphere are required. The models must
accommodate the extremes that the Sun may have encountered over
time.  For most of the discussion below we use a heliospheric
multifluid model that self-consistently incorporates the
charge-exchange interactions between interstellar neutral atoms
and the solar wind.  The dynamics of the solar wind, and therefore
the heliosphere characteristics, are altered substantially by
mass-loading from interstellar H$^\circ$ atoms, so modeling both
the pristine and charge-exchange H is a key aspect of
understanding variations in the heliosphere as the Sun passes
through different galactic environments.  These models form the
basis for the discussions in Sections 2 and 4.  For the extreme
and short-lived interstellar environment experienced by the
passage of an interstellar shock over the heliosphere, discussed
in Section 3, we use a different model.

\section{Basic Model Results\label{sec:models}}

To investigate the influence of the interstellar conditions on the
heliosphere, a detailed global heliospheric multifluid code is
used that self-consistently calculates plasma components and
neutral hydrogen \citep{Pauls95,Zank96,Mueller06}. Its strength
lies in the detailed treatment of the neutral component, which
originates in the ISM but is out of equilibrium when passing
through the heliosphere. This is the consequence of charge
exchange where neutral atoms are lost to the plasma, and new
neutrals are inserted into the distribution with velocity
characteristics that represent the underlying plasma protons.
Because the mean free paths can be quite long, the multifluid
approach treats the neutrals with three or four fluids, each
representing the characteristic major different plasma regions
where the respective secondary neutrals were born
\citep[e.g.,][]{Mueller08}.

We first establish a detailed global heliospheric multifluid code
that can accommodate the variable heliosphere boundary conditions
that the Sun may have encountered over time.  It assumes
axisymmetry about the stagnation axis (the axis through the Sun
that is parallel to the ISM flow) and an isotropic solar wind. The
model grid is polar, with an inner inflow boundary at 1 AU (or
even 0.7 AU for the smaller models of sections 2.3 and 2.4).
There, the 1 AU solar wind parameters of 5.0 \cc\ for the plasma
density, a temperature of $10^5$ K, and a radial velocity of 400
\kms\ are used as standard values; they represent a typical
in-ecliptic wind during solar minimum (slow solar wind). Each of
the neutral fluids interacts with the plasma through resonant \ce,
using the \citet{Fite} cross section, and all neutrals are
subjected to photoionization which depends on the squared distance
to the Sun. For simplicity, radiation pressure is assumed to
balance gravity for neutral hydrogen, and heliospheric and
interstellar magnetic fields are neglected. The interstellar
medium is prescribed as inflow boundary condition at a suitably
large distance from the interstellar bow shock (outer grid
boundary at 500 AU for section 2.3, 800 AU for section 2.4, 1000
AU for section 2.1, and 1500 AU for section 2.2). The boundary
parameters are the CHISM \HI\ and \HII\ number densities, and the
(common) hydrogen velocity and temperature. Outflow boundary
conditions are imposed on all fluids at the downwind outer
boundary, and at the inner boundary for all neutral fluids.
Photoionization already depletes the neutral density considerably
there. Secondary neutrals are permitted to escape also through an
outer outflow boundary condition in upwind directions.

\subsection{Contemporary ISM (LIC)\label{ss:contemporary}}

\begin{figure}
   \centering
   \includegraphics[bb=0 0 739 491,width=0.9\textwidth]{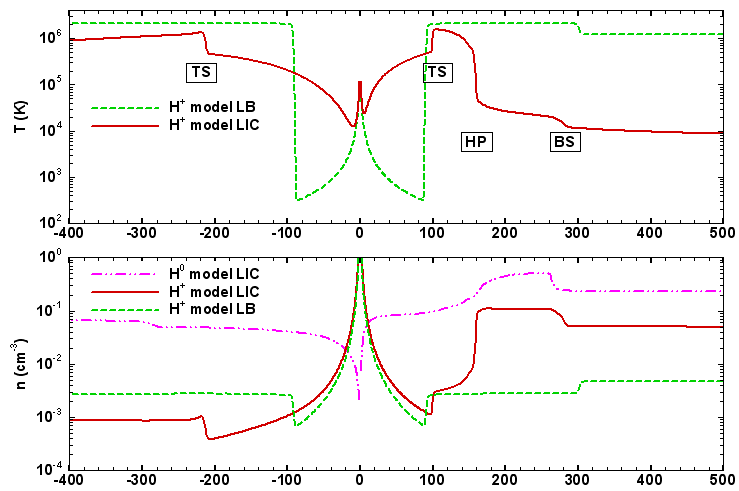}
      \caption{One-dimensional
profiles along the stagnation axis (distance in AU), with the Sun
at center (0) and the CHISM coming from right. Top: plasma
temperature of the Local Bubble case (dashed) and the contemporary
conditions (solid). The heliospheric boundaries of the LIC model
are marked in the plot. The bottom panel contains the
corresponding densities (LB plasma model, dashed; LIC model:
plasma, solid; neutral H, dash-dot pattern).}
         \label{fig:profile}
   \end{figure}

As a proxy for the contemporary values of the CHISM boundary
parameters we choose a model with \np\ = 0.047 \cc, \nH\ = 0.216
\cc, $v$ = 26 \kms, and $T$ = 7000~K \citep{SlavinFrisch:2008}.
Figure \ref{fig:profile} shows the plasma temperature (top panel)
and density (bottom panel) along the stagnation axis as solid
lines, together with the neutral H density (dash-dotted line). The
heliospheric boundaries appear as discontinuities. The
interstellar plasma goes through a bow shock (BS); the
accompanying decreased speed communicates via \ce\ to the neutrals
and creates a hydrogen wall \citep[hydrogen overdensity;][]
{Baranov93,LinskyWood96} in the post-bow shock region. On the
sunward side of the heliopause, the supersonic solar wind
undergoes a termination shock (TS) transition, and the hot
heliosheath plasma is diverted tailwards. Solar and interstellar
plasma are separated by a contact discontinuity, the heliopause
(HP). The thermodynamically distinct plasma regions (supersonic
solar wind; hot heliosheath; interstellar plasma) define the
characteristic heliospheric regions that form the basis of the
multifluid treatment of the neutrals in the model.

Figure \ref{fig:profile} also displays the temperature and density
profile of a plasma-only model, corresponding to the Local Bubble
environment (``LB'', dashed lines; presented in detail in section
\ref{ss:locbub} below). It demonstrates that \ce\ is the source of
solar wind heating from about 10 AU all the way to the TS: The
``LIC''-model plasma temperature is monotonically increasing in this
range, while the plasma-only model follows a strict adiabatic
cooling. At the HP, \ce\ organizes an anomalous heat transport from
solar heliosheath plasma to interstellar plasma; such an energy
transport is absent in the ``LB'' case without neutrals (HP at 300 AU
in Figure \ref{fig:profile}).

%-------------------------------------------------------------

The neutral atom density at the upwind TS is 0.098 \cc\ = 0.46
\nH. The factor 0.46 is called filtration factor, based on the
image that the pristine interstellar neutral flow gets processed
(``filtered'') while traversing the heliospheric interface between
BS and TS \citep{Wallis71}. Close to the Sun, photoionization and
increased charge exchange probability (due to increased solar wind
density) deplete neutral H exponentially, creating a neutral H
cavity. In the tail direction, the stagnation axis re-populates
slowly with off-axis neutral H.

\subsection{Hot Local Bubble\label{ss:locbub}}

The relative motions of the Sun and surrounding interstellar gas
indicate the Sun has emerged from the deepest void of the Local
Bubble interior within the past $\sim 130,000$ years.  Such
regions are common in the Milky Way Galaxy.  Following this
interpretation, the interior of this so-called Local Bubble is
assumed hot and highly ionized but of low density. We adopt the
ISM parameters of \np\ = 0.005 \cc, \nH\ = 0, $v$ = 13.4 \kms, and
log T(K) = 6.1 as a proxy model ``LB'' for such an interstellar
environment, with the velocity based on the
\citet{DehnenBinney:1998} solar apex motion since the LB plasma is
assumed to be at rest in the Local Standard of Rest (LSR). The
realization that part of the soft X-ray emission could be due to
heliospheric foreground composed of line emission of solar wind
charge exchange products being left in excited states, has
recently cast doubt whether the Local Bubble gas is in fact as hot
as stated above, so that the LB model outlined here can be taken
as an upper limit.

\begin{figure}
   \centering
   \includegraphics[bb=0 0 676 254,width=\textwidth]{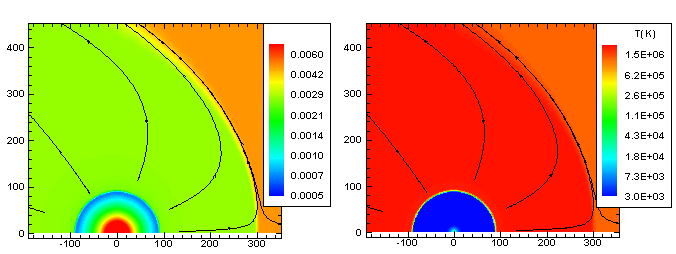}
      \caption{Plasma density (left) and temperature (right) for the heliosphere surrounded by Local Bubble gas.
              }
         \label{fig:LB}
   \end{figure}

The speed of sound in such a LB plasma is high, so that the solar
movement through it is decidedly subsonic. In this case, there is
no bow shock, and interstellar plasma gets decelerated
adiabatically to match the zero velocity of the stagnation point.
The isotropic thermal interstellar pressure dominates the ram
pressure, and the termination shock is spherical at a distance of
90 AU from the Sun. This distance is comparable to that of the
contemporary heliosphere. The distance to the nose of the
heliopause is 300 AU, which makes the heliosheath very large in
comparison to the one of the contemporary heliosphere. The
temperature in the sheath reaches values as high as $2.2\times
10^6$ K. Figure \ref{fig:LB} shows 2D maps of density and
temperature for the LB case, while Figure \ref{fig:profile}
contains both variables along the stagnation axis, as dashed
lines. In the termination shock transition, the density jumps by a
factor of 3.8, and the wind speed decreases to 100 \kms.

%-------------------------------------------------------------

In the LB model there are no neutral atoms in the entire combined
system of the solar wind and ISM.  The result is that a host of
physical effects present if neutral gas surrounds the Sun were
missing when the Sun was in the LB: There are no pickup ions (PUI)
produced by \ce, there are no anomalous cosmic rays, and no
slowdown or heating in the supersonic solar wind beyond the inner
solar system takes place. As mentioned by \citet{Mueller06}, the
plasma-only LB model exhibits a morphology and flow field that are
similar to the gross properties of the hydrodynamic problem of a
flow around a rigid sphere (excluding the flow detachment and
vortices occurring in the latter case). They also point to the
fact that absent the mitigating effects of neutrals, the magnetic
field of both solar and interstellar origin takes on a more
important role in the pressure balance, and corrections to the
above results can be expected when magnetic fields are included
realistically.

\subsection{Dense Neutral ISM\label{ss:hidens}}

\citet{Mueller06} have presented results from models with an ISM
density that is about two orders of magnitude higher than the
contemporary value. The resultant heliospheres are small because
of the increased interstellar ram pressure shifting the pressure
balance. The presence of a higher neutral density makes \ce\ a
more frequent occurrence, so that the plasma slowdown at the BS
gets immediately communicated to the neutrals with a resulting
hydrogen wall that is sharply defined post-BS, and has a high
amplitude. Although filtration is very effective, enough neutral
hydrogen enters the inner heliosphere inside the TS to lead to a
pronounced solar wind slowdown (values down to 260 \kms\ upstream
of the TS) and heating; the TS itself is therefore quite weak even
in the nose direction. The BS compression ratio, on the other
hand, is somewhat elevated. Atypically, the HP is not a sharp
temperature transition as in most other models, but the
temperature profile is more gradual, with the solar wind cooled
close to the HP by frequent \ce, and the interstellar plasma
heated close to the HP.

Even denser clouds encountering the heliosphere have been treated
by \citet{Yeghikyan03}, with $n_{\rm tot}$ = 100 \cc\ and higher.
Statistically such clouds could be encountered once every $\sim$10
Myrs, once per Gyr for giant molecular clouds (cold, 10K, with up
to 1000 \cc\ density). Not only does the heliosphere become
extremely small for the duration of the encounter, such that most
of the Earth orbit is in the ISM, but the presence of massive
amounts of neutral hydrogen makes it necessary to include
additional physics. For example, the interplanetary medium will no
longer by optically thin, as is a good assumption for contemporary
conditions. One important terrestrial consequence is that the high
density leads to removal of terrestrial atmospheric oxygen, and
other atmospheric effects \citep{Yeghikyan04terr}.

\subsection{High Velocity ISM\label{ss:hivel}}

Similarly to density, it is also worthwhile to explore variations
in the relative Sun-ISM speed. The interstellar velocity $V$ is a
key variable in understanding the ISM-heliosphere interactions
because the ram pressure varies as $V^2$.  As noted above, if the
ISM is at rest with respect to the Local Standard of Rest,
relative velocities of about 13 \kms\ result from the motion of
the Sun in that frame of reference. In the contemporary case, the
LIC moves with respect to the LSR, resulting in an overall 26
\kms\ relative motion. In addition, warm and cold \HI\ clouds have
radial velocities that vary from -80 to +6 km s$^{-1}$, with the
possibility that 3D velocities are even larger. For the purpose of
calculating astrospheres around stars in the solar neighborhood
that are still embedded in a partially ionized ISM,
\citet{Wood03,Wood05b} determine relative Star-ISM velocities for
selected objects. In their list there are entries with 68 and 86
\kms, and 40 Eri has a relative motion of 127 \kms\
\citep{Wood03}. When applying such relative motions to the Sun,
the resulting heliospheres can be described as ``wind-swept,''
i.e., a narrow leading cavity and a long, drawn-out tail
\citep{Mueller06}. The high velocity generates a large ram
pressure, making the resulting heliosphere smaller, similar to,
but more elongated than, the high density cases. Such heliospheres
also tend to have a triple point at the heliotail termination
shock, necessitated by the heliosheath flow accelerating to
supersonic velocities \citep[e.g.,][]{Pauls96}.

Not only are the high-speed heliospheres quite asymmetric, but the
neutral mean free paths are now on the order of the heliosheath
thickness or larger, meaning that the hydrogen wall is not very
pronounced, and neutral filtration is very weak so that the
interstellar density is not much higher than the neutral density
entering the inner heliosphere through the termination shock. Such
filtration ratios close to 1 seem only possible when a high
interstellar velocity combines with a modest or low density so
that the peak hydrogen wall occurs close to the HP without room
for depletion of neutral H between peak and HP.

\section{Supernova Remnant Encounters with the Heliosphere}\label{sec:snr}

The possibility that spikes in $^{10}$Be isotopes in the Antarctic
ice core samples were caused by a cosmic-ray enhancement due to a
supernova shock passing over the heliosphere was evaluated by
\cite{SonettJokipii:1987}. Initial supernova remnant (SNR)
expansion velocities are much larger than the high-velocity cases
discussed in section \ref{ss:hivel}. \citet{Fields08} have
recently calculated the effect on the heliosphere of a supernova
going off in the solar neighborhood. One of the primary
motivations of such studies is the evidence of live radioisotopes
in thin layers of ocean floors with lifetimes shorter than the age
of the solar system \citep{Knie99,Knie04}. A plausible explanation
of such data is that a SNR collision with Earth has deposited this
material ($^{60}$Fe) in a short-duration event $(2.8\pm 0.4)$ Myr
ago.

% Full-column wide figure: Supernova
\begin{figure*}
  \centering
  \includegraphics[bb=0 0 239 179,width=0.48\textwidth]{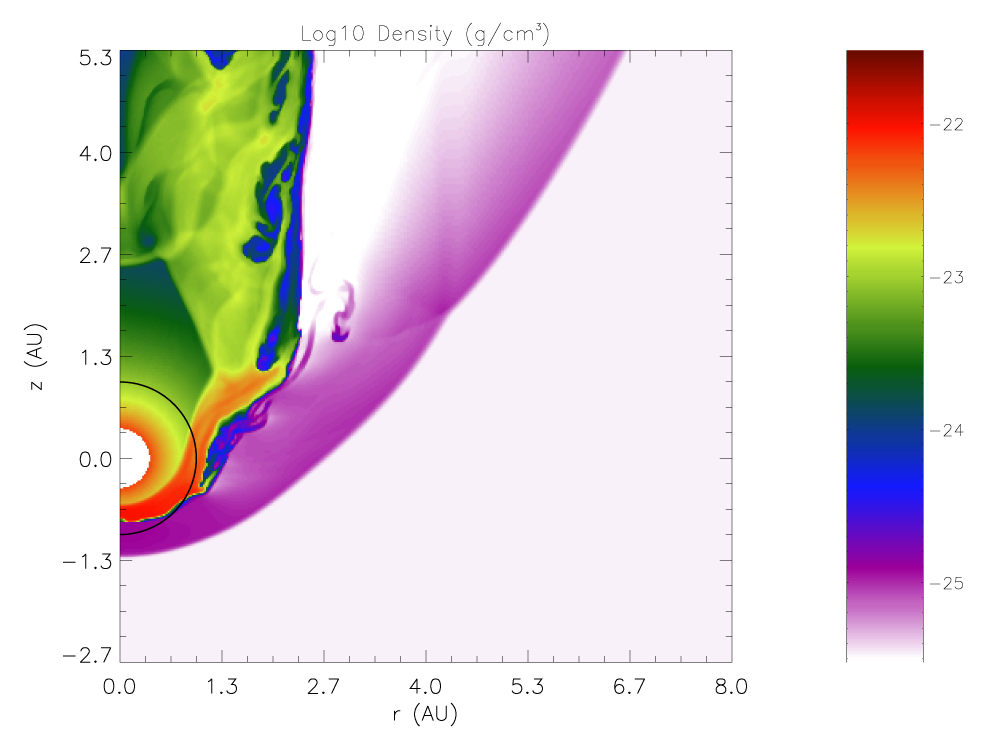}
  \includegraphics[bb=0 0 236 174,width=0.48\textwidth]{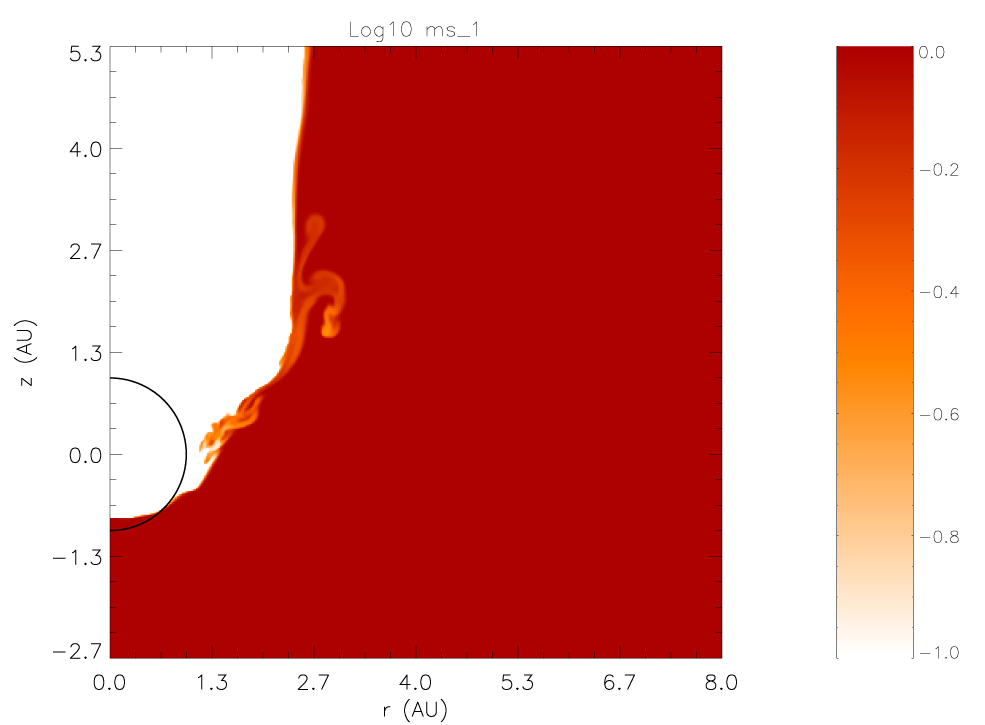}
\caption{(Left) Snapshot (logarithmic density map) of the
heliosphere during arrival of a shock from a supernova explosion 8
pc away, for LIC-like ambient ISM conditions (the SNR velocity is
12,000 \kms). A circle is drawn at a radius of 1 AU indicating the
Earth's orbit (modulo inclination effects). For a period of slow
solar wind, a SNR under these conditions will engulf the Earth,
and the Earth's orbit may carry it into the remnant where it will
be directly exposed to supernova material. (Right) Corresponding
contamination map (C = 1 supernova material, C = 0 solar
material). Supernova material penetrates to 1 AU; solar wind and
SNR plasmas mix also through instabilities at the heliopause.}
\label{fig:sn}       % Give a unique label
\end{figure*}

The effects of a SNR collision with the heliosphere depend on many
factors, starting with the initial explosive energy, the density
of the interstellar medium into which the supernova explodes, and
the supernova distance. SNR shock front density, pressure, and
velocity decrease with increasing distance from the supernova.
\citet{Fields08} have investigated a variety of ISM density
parameters and distances, and modeled the SNR evolution from
source to heliosphere with an AMR numerical code
\citep[FLASH;][]{Fryxell00}; neutrals are neglected.  Figure
\ref{fig:sn} displays an example of a heliosphere encountering a
SNR from a 10$^{51}$ erg supernova explosion at a distance of 8
pc, where the intervening ISM was assumed to have densities
similar to the Local Bubble. This distance corresponds to the
recent and conservative \citet{Gehrels03} estimate for the limit
at which the ionizing radiation from a supernova (prompt UV, X-
and gamma-ray photons, as well as the later diffusive cosmic rays)
inflicts damage on terrestrial stratospheric ozone at a level
which can cause severe damage to the biosphere.

The passage of the supernova blast profile persists for thousands
of years, with a slow secular decrease in pressure, velocity, and
density.  Consequently, high-speed, dense material comprises an
interstellar wind on time scales of many solar cycles, such that
on scales of the heliosphere the SNR blast is a basically steady,
plane-parallel ISM wind. The snapshot in Figure \ref{fig:sn} is
taken with solar wind conditions reflecting solar minimum. It can
be seen that the heliosphere is severely compressed, and the Earth
orbit (black circle) dips into the shocked interstellar medium for
part of the year, providing an interplanetary medium rich in
supernova ejecta that might eventually precipitate to Earth to
form the mentioned ocean bottom sediment layer.  We see that an
event close enough to cause biological damage is also able to
deliver supernova debris to the Earth, raising the possibility
that isotopic signatures can be correlated with possible
supernova-induced mass extinctions.

%It can also be seen that because of the prevailing high
%interstellar speeds, there is a lot of heliopause instability, and
%there is a high level of tail turbulence, mixing interstellar and
%solar wind material in the tail region (Figure \ref{fig:sn}b)

\section{Sensitivity of the Heliosphere to ISM Conditions}\label{sec:para}

Utilizing a broad set of plausible interstellar boundary
conditions for the heliosphere, it is possible to establish some
correlations between such parameters and the resultant heliosphere
configuration. To establish these correlations, numerous
multi-fluid models of the heliosphere and astrospheres have been
analyzed together. The models span the low-to-high interstellar
velocities and densities of \citet{Mueller06}, with most of them
obeying $500 < P/k < 6000$ \cc K, and a study with values probing
parameter space around the contemporary CHISM parameters with a
similar range of $P/k$ (see \citet{Mueller06i} for preliminary
results).
Regarding the latter study, the inferred values of the
contemporary CHISM are $v = 26.3$ \kms\ and $T \sim 6300 \pm 340$
K \citep{Witte96,Moebius04,Witte04}. The contemporary interstellar
proton and neutral H densities are not well constrained but should
lie in the range from 0.04--0.14 \cc\ and 0.14--0.24 \cc,
respectively \citep[e.g.][]{Zank99,SlavinFrisch:2002}.  The setup
for the systematic parameter study was therefore to probe the
distinct densities of 0.05, 0.11, 0.17, and 0.23 \cc\ for both the
interstellar neutral and plasma densities (16 combinations). The
CHISM ionization fraction ranges from 18\% to 82\%. To explore the
effects of temperature, it was decided to probe four temperatures
4000, 6000, 8000, and 10000 K, for a total of 64 models. The
velocity in this parameter study was set to 26.24 \kms.

One result from the parameter study as well as from the extension
to a wider parameter space is the predictability of the distance
of the upwind heliopause (stagnation point) as an expression of
the pressure balance between interstellar and solar wind. The HP
distance is calculated from the Rankine-Hugoniot termination shock
transition conditions and treating the heliosheath and shocked
interstellar flows as incompressible
\citep{SuessNerney90,Zank99,Mueller06}. A relation is obtained
that links the solar wind ram pressure $P_{1} = \rho_1^{\ }
v_{SW}^2$ at 1 AU and its scaling with heliocentric distance, to
the interstellar total pressure $P_{ISM}$. In a supersonic case,
the latter is dominated by $\rho_{ISM}^{\ } v_{ISM}^2$  with
$\rho_{ISM} = m_p [n({\rm HI}) + n({\rm HII})]$. The resulting
heliopause distance takes the form
\begin{equation}\label{eq-TS2}
r_{HP} = r_0\,\,
 \sqrt{\rho_1^{\ } v_{SW}^2\over P_{ISM}}
        \left( 1 - {v_{ISM}^2\over v_{SW}^2} \right) ^{5\over 4}\, .\\
\end{equation}
The constant $r_0$ is a product of factors from the theoretical
calculation and from the empirical fact that neutral hydrogen does
not fully participate in the pressure balance, but only weakly
couples to the plasma through \ce. A fit between model results and
$r_{HP}$ values yields a value of $r_0 = 1.70$ AU, with about a
6\% accuracy. A similar distance law was obeyed even by the
high-velocity models of \citet{Fields08} which represent SNR
collisions with the heliosphere (section 3). \citet{Baranov79}
have studied analytically the plasma-only case without neutrals
and arrive at a distance relation similar to equation
(\ref{eq-TS2}), without the last factor (typically close to unity)
involving the velocity ratio. Because of the entire ram pressure
$P_{ISM} = \rho_{p, ISM}^{\ } v_{ISM}^2$ participating fully in
the pressure balance in this case, they arrive at a lower $r_0 =
1.4$ AU, and the plasma-only heliospheric boundaries scale
self-similarly with the factor $(\rho_1^{\ } v_{SW}^2/ \rho_{p,
ISM}^{\ } v_{ISM}^2)^{1/2}$.

The magnetic fields of solar and interstellar origin are neglected
here; models that include interplanetary and interstellar magnetic
fields are presented elsewhere (see for example the Opher
contribution in this volume). It can be reasonably expected that
for models with neutrals and magnetic fields, the inclusion of
magnetic pressure in the pressure balance (\ref{eq-TS2}) will
preserve its validity to some extent. However, most sensible
interstellar magnetic field configurations render the heliosphere
asymmetric, and the distances of the boundaries as a function of
direction become more complicated than equations
(\ref{eq-TS2})-(\ref{eq-tsbscorr}) suggest.

In all available (non-magnetic) models, the nose distance of the
heliopause is accurately coupled to the termination shock
distance,
\begin{equation}
r_{\rm HP} = (1.40\pm 0.03) \,\, r_{\rm TS} .\label{eq-hpts}
\end{equation}
Again, this relation holds for SNR-type velocities as well
\citep[][Figure 7]{Fields08}. For the contemporary heliosphere,
this result means that if the termination shock lies between 86
and 94 AU, the heliosheath in upwind direction is between 35 and
38 AU thick, for a heliopause distance between 121 and 132 AU.

The interstellar bow shock distance also correlates with the
termination shock distance; however, there is an additional
temperature dependence. For a higher interstellar temperature, the
Mach number of the interstellar flow is lower, and hence the bow
shock itself weaker. To still achieve a zero velocity at the
stagnation point at the heliopause, the bow shock hence must be
further out for higher temperatures, in spite of the higher
interstellar pressure contributing to a slight shrinking of the
overall heliospheric size.  Multiplying a linear correlation with
an ad-hoc term that grows for increasing temperatures gives a very
good correlation between termination shock and bow shock distance,
\begin{equation}
r_{\rm BS} = 1.5 \,\, r_{TS}\,\,\left( 1+{1\over M-1} \right) + 23
{\rm AU}\,  \label{eq-tsbscorr}
\end{equation}
\citep{Mueller06i}.

The shape of the termination shock is not spherical, but it is
elongated in the tail directions. Typical values of the
downwind/upwind ratio range between 2 and 3. The tail TS distance
correlates with the upwind TS distance; however, there is again a
temperature influence in this correlation. The higher the
interstellar temperature, the lower the asymmetry of the TS
between downwind and upwind direction. Empirically there is a
correlation between $r_{TS}(180^{\circ})$ and the quantity
$r_{TS}(0^{\circ}) \sqrt{M}$, with a weak dependence of the
proportionality factor on the interstellar neutral hydrogen
density.

These results have been obtained from models where neutral
hydrogen was present and influenced the plasma via \ce. The
neutral hydrogen reaction to the different interstellar
environments resists categorization more than is the case for the
plasma results. For the 64-model systematic study, the height of
the hydrogen wall (the peak neutral density between BS and HP)
correlates with the interstellar Mach number. The higher the Mach
number, the higher the resulting hydrogen wall. An expected
anticorrelation between hydrogen wall height and amount of neutral
hydrogen entering through the termination shock into the inner
heliosphere could not be confirmed. There is, however, an
anticorrelation of this filtration factor $f$ with the
interstellar plasma density, $f \propto 1/\sqrt[3]{n({\rm HII})}$.

\section{Paleo-Heliosphere}\label{sec:paleo}

\begin{figure}
   \centering
   \includegraphics[bb=0 0 2102 1462,width=\textwidth]{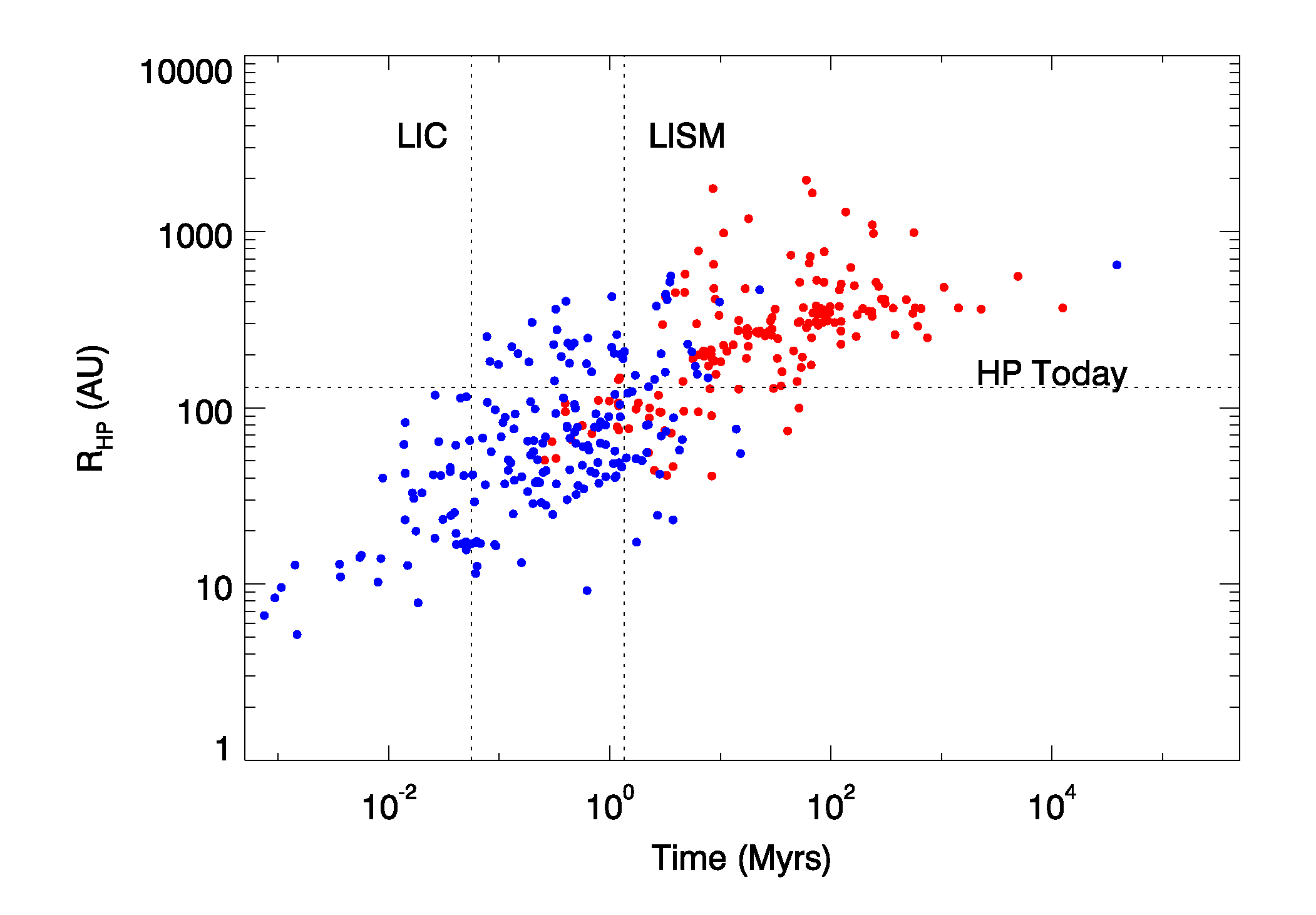}
      \caption{The heliosphere in time, according to the tropical
skies.  The length of time, in Myrs, that the Sun would spend in
the cold neutral (CNM) and warm neutral or partially ionized (WNM)
interstellar clouds detected in the Arecibo Millennium Survey is
plotted against the heliosphere radius predicted by eq.\ (1) for
those clouds. The CNM clouds (blue dots) assume densities of 15
cm$^{-3}$, and the WNM clouds (red dots) assume densities of 0.27
cm$^{-3}$. The length of times that the Sun spends in the LIC, and
in the very local interstellar material (LISM) is indicated, as is
the present heliopause distance in the upwind.  Since the true
three-dimensional velocity of these clouds is unknown, this
comparison assumes that the relative Sun-cloud velocity is the LSR
velocity of each component detected by Arecibo.  This plot shows
that if the Arecibo sky samples ISM that is typical of that
encountered by the Sun, then the Earth and Sun spends 99.4\% of
the time in WNM clouds.  }
         \label{fig:time}
   \end{figure}

The Arecibo Millennium Survey of the radio \HI\ sky \citep{HTIV}
provides reliable statistics on the distribution of the column
densities and velocities of warm and cold interstellar clouds seen
from the tropical sky.  About 60\% of interstellar \HI\ is warm
neutral material (WNM).  The median column density of the WNM,
$1.3 \times 10^{20}$ cm$^{-2}$, exceeds that of CNM, $0.5 \times
10^{20}$ cm$^{-2}$ \citep{Frisch08}. Although WNM is less dense
than CNM, higher typical WNM velocities and column densities
indicate that the Sun will spend more time in the WNM than the
CNM.  We calculate the length of time that the Sun would be in
each \HI\ component detected in the Arecibo survey, and compare
that time with the heliopause radius calculated from equation (1).
We use assumed densities of 0.27 \cc\ and 15 \cc\ for the WNM and
CNM, respectively, and assume that the observed LSR radial
velocity for each component is typical of the encounter velocity.
These values then give us a good estimate of the total length of
time the Sun is likely to spend in WNM versus CNM interstellar
gas. From Fig. \ref{fig:time} we quickly see that the Sun spends
significantly more time in WNM than CNM types of clouds. According
to the tropical Arecibo sky, the Sun is in WNM 99.4\% of the time,
but this estimate does not include the times spent in fully
ionized regions.   The Sun has entered the LIC within the past
$\sim 0.056$ Myrs, and the LISM within the past $\sim 0.13$ Myrs,
according to UV absorption lines towards nearby stars
\citep{FrischSlavin:2006}.

%-------------------------------------------------------------
%15. cm$^{-3}$

\section{Conclusions}\label{sec:last}

The ISM is a very dynamic environment. During its 5 Gyrs galactic
trajectory, the solar system likely has been embedded in a wide
variety of different interstellar environments. Some of these
environments lead to a drastically reduced heliosphere size,
allowing direct access of interstellar material to the solar
system and Earth. Even without direct ISM access, the particle
flux background like cosmic rays and dust is sensitive to the ISM
environment. The passage of the solar system through an arm of the
galaxy likely triggers a pronounced increase of both the rate of
supernovae going off near the Sun, and the flux of galactic cosmic
rays. Consequences for the terrestrial atmosphere/climate, and
their geological records, are likely
\citep[e.g.,][]{Frischetal:2006}, but not part of this paper.

When calculating the heliosphere while including the interstellar
neutrals self-consistent\-ly, and probing the vast interstellar
parameter space, it can be confirmed that the overall heliospheric
size is set by a pressure balance between solar wind and
interstellar medium, with only weak modifications involving the
interstellar Mach number and velocity (and most likely the
magnetic pressure which was not included in this present
analysis). The termination shock, heliopause, bow shock, and
downwind termination shock locations exhibit simple correlations
to the location of pressure balance. Neutral results like the
filtration by the heliospheric interface, and the height of the
hydrogen wall, also show a dependence on the interstellar
parameters. For an encounter with dense ISM or very fast ISM,
particle fluxes like neutrals, cosmic rays, and dust increase
markedly at Earth orbit. A passage of a remnant of a nearby
supernova blast has the ability to compress the heliosphere to a
size that allows interstellar material including supernova ejecta
direct access to Earth. In contrast, during the passage of the Sun
through the Local Bubble, or through any almost completely ionized
region, neutrals, pickup ions, and anomalous cosmic rays were
entirely absent.

\begin{acknowledgements}
HRM and PCF thank the International Space Science Institute
(ISSI), Bern, Switzerland, for hosting the 2007 workshop ``From
the Heliosphere to the Local Bubble'' and partially funding
participation.  HRM acknowledges partial funding through NSF grant
AST-0607641 and NASA SHP grants NNG06GD48G and NNG06GD55G. PCF
acknowledges funding through NASA grants NAG5-13107 and
NNG05GD36G. The work of BDF  was supported by NASA Exobiology
grant EXB03--0000-0031.
\end{acknowledgements}
\newcommand{\bibfont}{\footnotesize}
\bibliographystyle{SSRv}
%\bibliography{hrm_refs}   % name your BibTeX data base

%% Non-BibTeX users please use
\hyphenation{Post-Script Sprin-ger}

\end{document}